\documentclass[twoside]{article}
\usepackage{fleqn,espcrc2,amssymb}
\usepackage[dvips]{graphicx}

\newcommand{\AmS}{{\protect\the\textfont2
  A\kern-.1667em\lower.5ex\hbox{M}\kern-.125emS}}
\title{$s$- and $d$-wave solution of Eliashberg equations with finite bandwidth}
\author{G.A.~Ummarino and R.S.~Gonnelli\\
\vspace{3mm} INFM - Dipartimento di Fisica, Politecnico di Torino,
c.so Duca degli Abruzzi 24, 10129 Torino, Italy}

\begin{document}

\begin{abstract}
In this work, we discuss the results of the direct solution of the
Eliashberg equations with finite bandwidth, in the cases of $s$-
and $d$-wave symmetry for the pair wave function and in the
presence of scattering from impurities. We show that the reduction
of the critical temperature $T_{\mathrm{c}}$ due to the finite
bandwidth depends on the value of the bandwidth itself, but is
almost independent of the symmetry
of the order parameter. The same happens for the shape of $Z(\omega%
)$ and $\Delta(\omega)$. Moreover, we discuss the effect of the
finite bandwidth on the shape of the quasiparticle density of
states. The results clearly indicate that the infinite bandwidth
approximation leads to an underestimation of the electron-boson
coupling constant.
\end{abstract}

\maketitle

High-$T_{\mathrm{c}}$ cuprates and fullerenes are characterized by
phononic energies ($\Omega _{\mathrm{phon}}$) comparable with the
electronic ones ($E_\mathrm{F}$) while in low-$T_{\mathrm{c}}$
superconductors it is always $E_{\mathrm{F}}\gg \Omega
_{\mathrm{phon}}$. This last condition leads to the standard
Eliashberg equations \cite {refe1} obtained within the limit
$E_{\mathrm{F}}\rightarrow +\infty$. In this paper, we study the
effect of a finite bandwidth \cite{refe8a,refe20} on some relevant
physical quantities in the framework of the Eliashberg theory. For
simplicity, we disregard the important related problem of the
breakdown of the Migdal's theorem, and put to zero the Coulomb
pseudopotential $\mu ^{*}$ \cite {refe8}. The kernels of the
Eliashberg
equations (EE) for the renormalization function $Z(\omega%
,{\mathbf k})$ and the order parameter $\Delta (\omega ,{\bf k})$
contain the retarded electron-boson interaction $\alpha^2(\Omega%
,{\bf k},{\bf k}^{\prime})F(\Omega )$. Referring to
high-$T_\mathrm{c}$ cuprates, we assume ${\mathbf k}$ and
${\mathbf k}^{\prime }$ to
lie in the $ab$ plane (CuO$_2$ plane) and call $\phi$ and $\phi%
^{\prime }$ their azimuthal angles in this plane
\cite{refe20,refe10,refe15}. We integrate over the effective band
normal to the Fermi surface from $-W$ to $+W$, and expand
$\alpha^{2}(\Omega ,\phi ,\phi ^{\prime })F(\Omega )$ in terms of
basis functions $\psi _0\left( \phi \right) =1$ and $\psi _1\left(
\phi \right) =\sqrt{2}\cos \left( 2\phi \right) $ in the following
way:
\begin{eqnarray}
\alpha ^2(\Omega ,\phi ,\phi ^{\prime })F(\Omega )&=&\alpha
_{00}^2F(\Omega )\psi _0\left( \phi \right) \psi _0\left( \phi
^{\prime }\right)\cr\nonumber
 & & \;+ \alpha _{11}^2F(\Omega )\psi _1\left( \phi
\right) \psi _1\left( \phi ^{\prime }\right).
\end{eqnarray}

We then search for a $s+\mathrm{i}d$ solution with
\begin{eqnarray}
\Delta _{n}(\phi )\!\!\!\!&\equiv\!\!\!\!&\Delta(\mathrm{i}\omega
_{n},\phi )=\Delta _{s}(\mathrm{i}\omega _{n})+\Delta
_{d}(\mathrm{i}\omega _{n})\psi _{1}\left( \phi \right)
\cr\nonumber Z_{n}(\phi )\!\!\!\!&\equiv\!\!\!\!&
Z(\mathrm{i}\omega _{n},\phi )= Z_{s}(\mathrm{i}\omega
_{n})+Z_{d}(\mathrm{i}\omega
_{n})\psi _{1}\left( \phi \right)\nonumber.%
\end{eqnarray}

In this case, and in the Matsubara representation, the Eliashberg
equations in the presence of impurities become
\cite{refe8a,refe20}:
\vspace{-2mm}
\[
\omega _{n}Z_{n}(\phi )=\omega _{n}+\frac{k_{B}T}{\pi
}\sum_{m=-\infty }^{+\infty }\int_{0}^{2\pi }\mathrm{d}\phi
^{\prime }
\]
\vspace{-6mm}
\[
\hspace{2mm} \left[\lambda_{m,n} \left(\phi,\phi ^{\prime
}\right)N_{m}(\phi^{\prime } )\theta _{m}(\phi^{\prime
})+\frac{N_{n}(\phi^{\prime } )\theta _{n}(\phi^{\prime }
)}{k_{B}T\tau } \right] \label{3}
\]
\[
\Delta_{n}(\phi )Z_{n}(\phi )=\frac{k_{B}T}{\pi }%
\sum_{m=-\infty }^{+\infty }\int_{0}^{2\pi }\mathrm{d}\phi
^{\prime}
\]
\vspace{-6mm}
\[
\hspace{2mm} \left[\lambda_{m,n}\left(\phi,\phi ^{\prime
}\right)P_{m}(\phi^{\prime } )\theta _{m}(\phi^{\prime }
)+\frac{P_{n}(\phi^{\prime } )\theta _{n}(\phi^{\prime }
)}{k_{B}T\tau }\right]
\]
\[
\theta _n(\phi )=\tan ^{-1}\left[ W/2\sqrt{\omega _n^2Z _n^2(\phi
)+\Delta _n^2(\phi )Z _n^2(\phi )}\right]\nonumber
\]
\[
\lambda_{m,n} \left(\phi,\phi^{\prime } \right)=2\int_{0}^{+\infty
}\mathrm{d}\Omega \frac{\Omega \alpha ^{2}(\Omega ,\phi ,\phi
^{\prime })F(\Omega )}{\Omega ^{2}+(\omega _{n}-\omega _{m})^{2}}
\]
where $\omega_n=\pi (2n+1)k_BT$, and $\tau ^{-1}$ is the impurity
scattering rate. Furthermore we know that:
\begin{eqnarray}
P_n(\phi )\!\!\!\!&=\!\!\!\!&\Delta _{n}(\phi )Z _{n}(\phi
)/\sqrt{\omega _n^2Z _{n}^2(\phi )+\Delta _{n}
^2(\phi)Z _{n}^2(\phi)}\cr\cr\nonumber%
N_n(\phi )\!\!\!\!&=\!\!\!\!&\omega_nZ _{n}(\phi)/\sqrt{\omega
_n^2Z _{n}^2(\phi )+\Delta _{n}^2(\phi )Z _{n}^2(\phi )}
\nonumber.
\end{eqnarray}
\vspace{-4mm}

Thus we have four equations for $Z_{s}(\mathrm{i}\omega_{n} )$,
$Z_{d}(\mathrm{i}\omega_{n} )$, $\Delta%
_{s}(\mathrm{i}\omega _{n})$ and $\Delta _{d}(\mathrm{i}\omega
_{n})$. Here we only consider the case $Z_{d}(\mathrm{i}\omega
_{n})\equiv 0$ \cite{refe15}.

In our numerical analysis we put, for simplicity, $\alpha
_{11}^2F(\Omega )$=$g_d\!\cdot\!\alpha _{00}^2F(\Omega )$ where
$g_d$ is a constant and, as a consequence, $\lambda
_d=g_d\!\cdot\!\lambda _s$ \cite{refe20,refe15}. We used for
$\alpha _{00}^2F(\Omega )$
the spectral function we experimentally determined in Bi$%
_2$Sr$_2$CaCu$_2$O$_8$ (BSCCO) \cite{refe18} appropriately scaled
to give, in both the $s$ and $d$ cases, $T_\mathrm{c}=97$ K in the
$W=\infty$ limit. The coupling constants result to be $\lambda%
_s=3.15$ for the $s$-wave case, $\lambda _s=2$ and $\lambda%
_d=2.3$ for the $d$-wave one.
\begin{figure}[t]\vspace{-2mm}
\includegraphics[keepaspectratio,width=7.5cm]{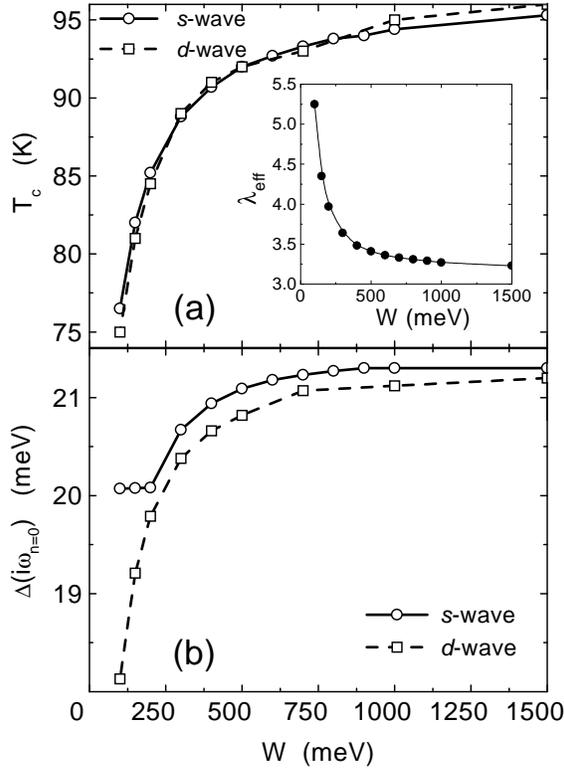}\vspace{-10mm}
\caption{ \small{(a) The critical temperature $T_\mathrm{c}$
versus the bandwidth $W$. (b) $\Delta(\mathrm{i}\omega _{n=0} )$,
at $T$=4 K, versus $W$. The inset of (a) reports $\lambda _{eff}$
versus $W$.} } \vspace{-8mm}
\end{figure}
From the direct numerical solution of the EE we found that the
symmetry of $\Delta_n (\phi)$ depends on the values of the
coupling constants $\lambda _s$ and $\lambda _d$ and, for
particular values of $\lambda _s$ and $\lambda _d$, on the
starting values of $\Delta _s$ and $\Delta _d$.

As shown in Fig.~1 (a) and (b), the dependence of $T_\mathrm{c}$
and $\Delta(\mathrm{i}\omega_{n=0})$ at $T$=4~K on the bandwidth
$W$ is almost the same in both the $s$- and $d$-wave symmetries.
The reduction of the bandwidth results in a sensible reduction of
$T_\mathrm{c}$ and, on a minor extent, of
$\Delta(\mathrm{i}\omega_{n=0})$. In Fig.~2 we show the $d$-wave
normalized density of states, calculated through the analytical
continuation with Pad\'e approximants of the imaginary-axis
solution, for various values of $W$ at $T$=2 K \cite{refe20}. As
in the $s$-wave case (not shown), the effect of a small bandwidth
is remarkable particularly at high energies. These curves are in
\emph{better agreement} with tunneling experimental data than
those obtained for an infinite bandwidth, particularly for the
presence of a \emph{dip} above the peak (at about $2\Delta$) and
because they asymptotically tend to 1 from above.

We have also calculated the real and imaginary parts of the
functions $\Delta (\omega )$ and $Z(\omega )$ in the $s$- and
$d$-wave cases. For small values of $W$, these functions are
markedly modified with respect to the infinite bandwidth case,
independently of the symmetry. In the inset of Fig.~1(a) we show
(only in the $s$-wave case) the values of $\lambda _\mathrm{eff}$
necessary to obtain $T_\mathrm{c}=97$ K, for various values of
$W$. We conclude that the use of the standard Eliahsberg equations
when $E_\mathrm{F}\simeq \Omega _{\mathrm{phon}}$ leads to
\emph{underestimate} the real value of the electron-boson coupling
constant.

\begin{figure}[t]
\includegraphics[keepaspectratio,width=7.5cm]{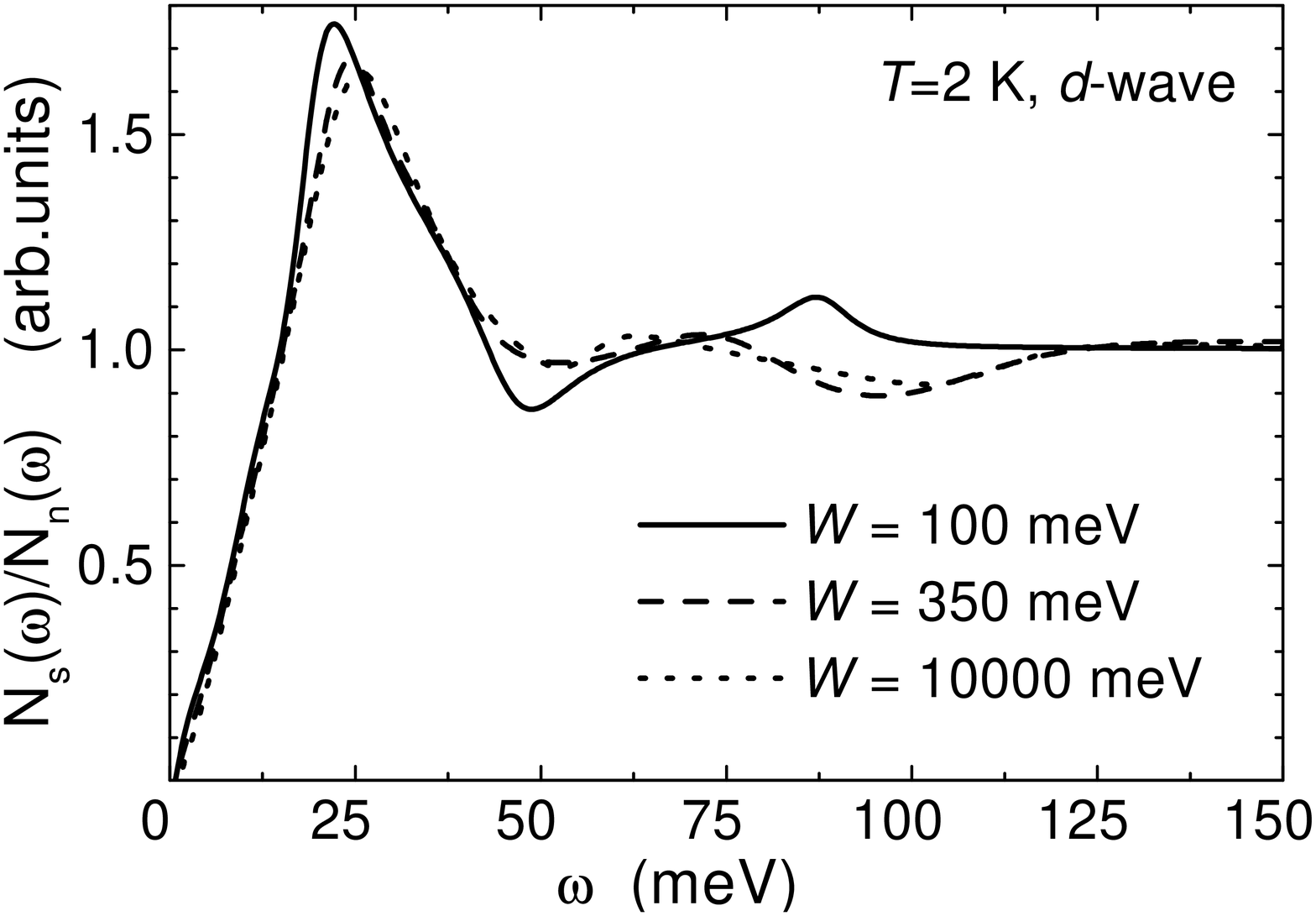}\vspace{-10mm}
\caption{
\small{The $d$-wave normalized density of states, at $T= 2$~K, %
for $W$=100, 350 and 10000 meV.} }
\vspace{-4mm}
\end{figure}

\small{

}
\end{document}